\documentclass{emulateapj}
\usepackage[T1]{fontenc}
\usepackage[latin9]{inputenc}
\usepackage{amsmath}
\usepackage{amssymb}
\usepackage{graphicx}
\usepackage{natbib}
\usepackage{multirow}
\usepackage{psfrag}
\usepackage{color}
\usepackage{ulem}

\newcommand{\msol}{M_\odot}
\newcommand{\rsol}{R_\odot}

\newcommand{\noun}[1]{\textsc{#1}}

\shorttitle{Eclipsing AM CVn System}
\shortauthors{Levitan et al.}

\begin{document}

\title{PTF1\,J191905.19+481506.2 --- A Partially Eclipsing AM CVn System Discovered in the Palomar Transient Factory}

\author{David Levitan\altaffilmark{1},  Thomas Kupfer\altaffilmark{2},
Paul J. Groot\altaffilmark{1,2},   Bruce Margon\altaffilmark{3}, 
Thomas A. Prince\altaffilmark{1}, Shrinivas R. Kulkarni\altaffilmark{1},
Gregg Hallinan\altaffilmark{1}, Leon K. Harding\altaffilmark{1}, Gillian Kyne\altaffilmark{4},
Russ Laher\altaffilmark{5},
Eran O. Ofek\altaffilmark{6},
Ren\'e G. M. Rutten\altaffilmark{7},
Branimir Sesar\altaffilmark{1},
 and Jason Surace\altaffilmark{5}}

\altaffiltext{1}{Division of Physics, Mathematics, and Astronomy, California Institute of Technology, Pasadena, CA 91125, USA.}
\altaffiltext{2}{Department of Astrophysics/IMAPP, Radboud University Nijmegen, PO Box 9010, NL-6500 GL Nijmegen, the Netherlands.}
\altaffiltext{3}{Department of Astronomy and Astrophysics, University of California, 1156 High Street, Santa Cruz, CA 95064, USA.}
\altaffiltext{4}{Centre for Astronomy, School of Physics, National University of Ireland, Galway, Ireland.}
\altaffiltext{5}{Spitzer Science Center, MS 314-6, California Institute of Technology, Pasadena, CA 91125, USA}
\altaffiltext{6}{Benoziyo Center for Astrophysics, Faculty of Physics, Weizmann Institute of Science, Rehovot 76100, Israel}
\altaffiltext{7}{GRANTECAN S.A. Cuesta de San Jos\'e, s/n - 38712 - Bre\~na Baja - La Palma, Espa\~na.}

\begin{abstract}
We report on PTF1\,J191905.19+481506.2, a newly discovered, partially eclipsing, outbursting AM CVn system found in the Palomar Transient Factory
synoptic survey. This is only the second known eclipsing AM CVn system. We use high-speed photometric observations and phase-resolved spectroscopy to establish an orbital period of $22.4559(3)\,$min.
We also present a long-term light curve and report on the normal and super-outbursts regularly seen in this system, including
a super-outburst recurrence time of $36.8(4)\,$d. We use the presence of the  eclipse to place upper and lower limits on the inclination of the system and
discuss the number of known eclipsing AM CVn systems versus what would be expected.
\end{abstract}

\keywords{accretion, accretion disks --- binaries: close --- novae, cataclysmic variables --- stars: individual: (PTF1\,J191905.19+481506.2) --- white dwarfs}

\section{Introduction}

AM CVn systems are rare, ultra-compact, semi-detached, white dwarf binaries with periods ranging from 5 to
65 minutes. First identified over 40 years ago by \citet{1967AcA....17..255S}, it has only been
in the last decade that the number of known systems has risen above ten. Yet despite the recent discovery
of almost 25 additional systems, their rich and complex phenomenological behavior
has limited our understanding of this class of post-period minimum binaries. They are considered to be the helium analog
of cataclysmic variables (CVs) and are important
as strong, low-frequency Galactic
gravitational wave sources \citep{2004MNRAS.349..181N,2007MNRAS.382..685R,Nissanke:2012fk},
the source of the proposed ``.Ia'' supernovae \citep{2007ApJ...662L..95B}, and one of the
believed end-points of binary white dwarf evolution \citep{2001A&A...368..939N}. 
However, many questions about these unique systems remain, including
their population density, their evolutionary pathways, and the interactions between the
two components, including the He-rich accretion disk. We refer the reader to \citet{2005ASPC..330...27N}
and \citet{2010PASP..122.1133S} for general reviews.

Follow-up study of a newly-identify AM CVn system typically begins with a spectroscopic determination of its orbital
period using phase-resolved spectroscopy of the ``hot spot'' \citep{1981ApJ...244..269N}, which is the location where matter
from the donor hits the accretion disk. However, these measurements are incomplete, as the inclination of the systems cannot be determined.
Only those systems showing eclipses and radial velocity variations of
two components (e.g., the donor and accretor or the accretor and hot spot) allow for a full determination of the system's
parameters (e.g., component masses, inclination, period, etc).
Eclipses also allow for the extremely precise measurement of the
orbital period change. For example, the orbital period of the double
detached white-dwarf binary SDSSJ0651+2844 was measured to be
decreasing by \citet{Hermes:2012ys} while the only known eclipsing AM
CVn system, SDSSJ0926+3624 (\citealt{2005AJ....130.2230A}; \citealt{2011MNRAS.410.1113C}, hereafter C11),
was recently shown to have an increasing orbital period \citep{2013arXiv1309.4087S}.

AM CVn systems have been observed to have three relatively distinct phenomenological states. 
Systems with orbital periods (hereafter $P_{orb}$) less than $20\,$min have been observed in a ``high'' state, characterized by optically
thick accretion disks and absorption-line spectra. In contrast, systems with $P_{orb}>40\,$min,
are observed to have emission-line spectra from what are believed to be optically-thin accretion
disks and are said to be in quiescence (though see \citealt{Woudt:2013kx} for an example
of a longer period system that has been observed to outburst). Systems in neither class show photometric variability over 0.5\,mag.

Between these orbital period limits are outbursting systems, which are observed to
have dwarf nova-type outbursts of 3--6\,mag as well as
variability at the 10\% level in both quiescence and outburst \citep[e.g., ][]{1997PASP..109.1100P}.
Recent studies have shown that the frequency of these outbursts decreases as the orbital period
increases (\citealt{2011ApJ...739...68L}; \citealt{2012MNRAS.419.2836R}).

The population density of AM CVn systems has not been conclusively determined:
population synthesis estimates of the space density \citep{2001A&A...368..939N} have not been
observationally confirmed by color-selected samples \citep{2007MNRAS.382..685R,Carter:2013fk}.
The latest results of \citet{Carter:2013fk} suggest a space density of $(5\pm3)\times10^{-7}\,\text{pc}^{-3}$,
a factor of 50 lower than the population synthesis estimates by \citet{2001A&A...368..939N}. The
reason for this discrepancy is currently unknown. We note that these color-selected samples
are mostly sensitive to longer period systems.

Over the last two years, we have conducted a search for outbursting AM CVn systems
using the Palomar Transient Factory\footnote{\url{http://ptf.caltech.edu/}}
\citep[PTF; ][]{2009PASP..121.1395L,2009PASP..121.1334R} large-area synoptic survey, in part, to use a different
approach to the discovery of AM CVn systems that does not rely on their colors.
The PTF uses the Palomar $48^{\prime\prime}$ Samuel Oschin Schmidt telescope
to image up to $\sim2,000\deg^{2}$ of the sky per night to a depth of $R\sim20.6$ or $g'\sim21.3$.
After identifying outbursting systems, we obtain classification spectra. To date, this survey
has identified $\mathord{>}340$ cataclysmic variables (CVs), six new AM CVn systems,
and one extremely faint AM CVn candidate \citep{2011ApJ...739...68L,Levitan:2013uq}.

In this paper, we present PTF1\,J191905.19+481506.2 --- a new AM CVn system discovered by
the Palomar Transient Factory. This system is particularly interesting because:
\begin{itemize}
\item shallow eclipses are present, making it only the second known eclipsing AM CVn system.
\item its orbital period is the shortest known of outbursting AM CVn systems.
\end{itemize}
We note that while this system is in the Kepler field, it, unfortunately, fell into a gap between the detectors
of the \textit{Kepler} satellite.

This paper is organized as follows. In Section \ref{sec:detection} we discuss our data reduction and analysis methods.
In Section \ref{sec:observations} we present both short-term and long-term photometric and spectroscopic data
and perform a period analysis. We consider the system's geometric structure
and the rate of eclipsing AM CVn systems in Section \ref{sec:discussion}.
We summarize in Section \ref{sec:conclusions}.

\section{Data Acquisition, Reduction, and Analysis}
\label{sec:detection}

\subsection{Photometric Data}
\label{sec:photomdata}

The initial discovery of PTF1\,J191905.19+481506.2, hereafter PTF1J1919+4815,
as an outbursting compact binary candidate was made using the PTF.
Two pipelines process PTF data. The ``transient'' pipeline uses difference imaging to identify possible 
transients in real-time \citep{Gal-Yam:2011ve}. In contrast, the
``photometric'' pipeline prioritizes photometric accuracy at the cost of processing time and uses aperture photometry
to measure fluxes. This paper uses data from the latter.

The photometric pipeline applies standard de-biasing, flat-fielding, and astrometric calibration to raw images
\citep{LAHERPTF}. Absolute photometric calibration to the few percent level is performed using a fit to SDSS fields observed in the same night
\citep{2012PASP..124...62O}. Additional relative photometric calibration is applied to improve precision to 6--8\,mmag at the bright end
of $R\sim14$ and 0.2\,mag  at the faint end of $R\sim20.6$. This algorithm is described, in part, in Section \ref{sec:relphot}.

While the initial identification of the system
as an outbursting, compact binary candidate was done using data from the aforementioned pipeline, the crowded nature
of the field requires point spread function (PSF) photometry for optimal results. We therefore re-processed the PTF images using the same pipeline
as described in Section \ref{sec:relphot}.

Dedicated long-term monitoring was obtained using the Palomar $60^{\prime\prime}$ (P60) telescope. 
The P60 automated pipeline, which includes automated de-biasing, flat-fielding, and astrometric calibration,
is described in \citet{2006PASP..118.1396C}.

High-cadence observations were obtained from two sources.
The first observations were made using the Lick 3-m Shane telescope with the Kast
imaging spectrograph in imaging mode. We inserted a $g'$ filter into the user filter wheel,
replaced the dichroic with a mirror, and used a clear window instead of a grism. This
provided an approximately $2'\times2'$ field of view in a single filter with a dead time of $3.0\,$s between exposures 
utilizing the fast CCD read-out mode and disabling auto-erase of the CCD.
Data were de-biased and flat-fielded using standard routines and astrometrically calibrated
using the \noun{starlink} package \noun{autoastrom}.

More recent high-cadence observations were obtained with the Caltech High speed Multi-color
camERA\footnote{for more information contact PI G.~H., instrument scientist L.~K.~H. or see \url{http://tauceti.caltech.edu/chimera/}}
(CHIMERA), recently developed for the Palomar $200^{\prime\prime}$ (5.1\,m) telescope's (P200)
prime focus. Incoming light
is split using a dichroic element onto two cameras, first passing through a $g'$ filter on one side
and an $r'$ or $i'$ filter on the other. 
The instrument can interchangeably make use of either Andor NEO sCMOS cameras or
Andor iXon 897 Ultra EM-CCD cameras, depending on the requirements for each program
(field of view, cadence, red versus blue response).

For this paper, we use only $g'$ data captured by an Andor EM-CCD. This camera, when installed in
CHIMERA, provides a $2^{\prime}\times2^{\prime}$ field of view with $0.35^{\prime\prime}\,\text{pixel}^{-1}$
plate scale. Exposures were de-biased and flat-fielded using standard routines. Photometric calibration
was performed as described in Section \ref{sec:relphot}, with the exception that the photometric measurements
were made using the \noun{apphot} package in \noun{iraf}.

\subsubsection{Photometric Calibration}
\label{sec:relphot}

Photometric measurements for the PTF, P60, and Lick data presented here were made using
PSF photometry as implemented by the \noun{autophotom} package. Photometric calibration was performed for all photometric data
using a least-squares matrix algorithm  described in \citet{2011ApJ...740...65O} and \citet{2011ApJ...739...68L}.
The algorithm is similar to that in \citet{1992PASP..104..435H}, but allows for a simultaneous fit
to reference magnitudes, providing both absolute and relative calibration in one step.
The absolute calibration for this data was derived from USNO B-1.0, which is
accurate to 0.5\,mag \citep{2006AJ....131.2801S}. We note
that the same algorithm is used for the photometric PTF pipeline, except that the photometric PTF pipeline
is based on photometric measurements from Sextractor \citep{1996A&AS..117..393B}.

\subsection{Spectroscopic Data}

Spectroscopic data were acquired from a number of telescopes and instruments; all were long-slit spectrographs. 
The observations are detailed in Section \ref{sec:observations}. The initial
identification spectra obtained using the P200 were reduced using standard IRAF routines. 
All follow-up phase-resolved spectroscopic data were reduced using optimal extraction \citep{1986PASP...98..609H}
as implemented in the \noun{Pamela} code \citep{1989PASP..101.1032M} as well
as the \noun{Starlink} packages \noun{kappa}, \noun{figaro}, and \noun{convert}. Spectra obtained from the red side of
Keck-I/LRIS were processed with \noun{L.A. Cosmic} \citep{2001PASP..113.1420V} due to the large number of cosmic rays.

\subsection{Period Estimation}
\label{sec:perest}
We use Lomb-Scargle periodograms (\citealt{1982ApJ...263..835S}; implementation by \citealt{2011ApJ...733...10R})  to identify periods in the data. Error estimation
is performed using a bootstrap approach \citep{1982jbor.book.....E}. For each data set of length $N$ points, we draw $N$ points at random, allowing for repetition.
We then generate a Lomb-Scargle periodogram for the selected points and find the peak. This method allows us to randomly vary both the
length of the data set and the points from which a period is calculated. We repeat this process 500 times, and take their robust standard
deviation, defined as $\sigma_\text{rob}=0.741(75^\text{th}~\text{percentile} - 25^\text{th}~\text{percentile})$, as an estimate of the error.

\section{Observations and Period Analysis}
\label{sec:observations}

\begin{figure}
\includegraphics{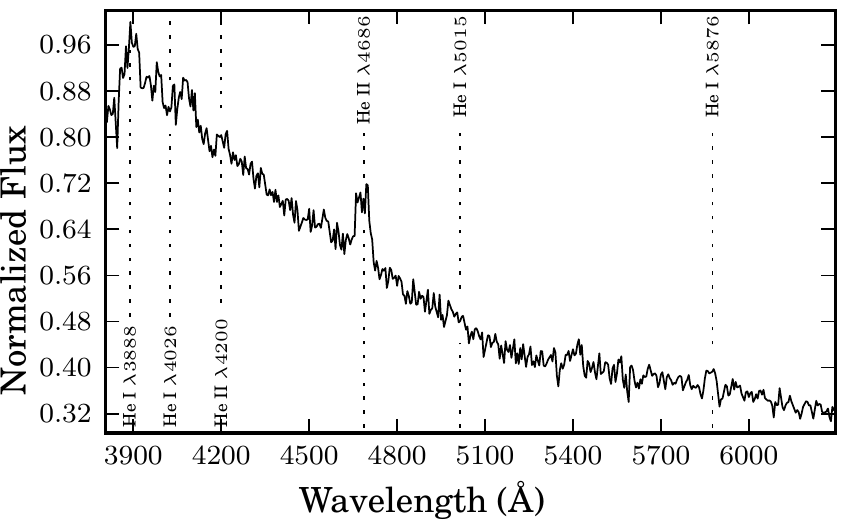}
\caption{The classification spectrum of PTF1J1919+4815 obtained using the DBSP instrument on the Palomar $200^{\prime\prime}$.
Significant lines are identified. The absence of Balmer-series lines and the presence of He lines indicated that this was a likely
AM CVn system. In contrast with the quiescent
spectra of AM CVn systems, very few \ion{He}{1} lines show significant emission, most notably \ion{He}{1} $\lambda5875$, 
while \ion{He}{2} $\lambda4686$ is very strong with an equivalent width of $-11.4\pm0.8$\AA.\label{fig:initspec}}
\end{figure}

PTF1J1919+4815 was detected as a possible transient on 2011 July 11. A spectrum obtained at the
Palomar $200^{\prime\prime}$ (P200) on
2011 July 23, likely while the system was still in outburst, showed a strong \ion{He}{2} 4686 emission line, but no other
significant spectral lines (Figure \ref{fig:initspec}). A second spectrum taken on 2011 August 03 detected the system
while it was in a much fainter state,
resulting in much lower signal-to-noise with few discernible lines. The
combination of the crowded field (relative to the PTF's pixel scale of $1.01^{\prime\prime}\,\text{pix}^{-1}$) and
lack of a clear spectral signature led to the object being left for future study.

In 2011 December, PTF1J1919+4815 was identified as a candidate outbursting source as part of the PTF search for outbursting
sources (\citealt{Levitan:2013uq}). The combination of no Balmer lines and the presence of \ion{He}{2} in the initial spectrum,
as well as the outbursting behavior, led to its initial classification as an AM CVn system candidate. The likelihood that PTF1J1919+4815 was a 
short-period system made phase-resolved observations necessary to understand its nature.

\begin{figure}
\begin{centering}
\includegraphics{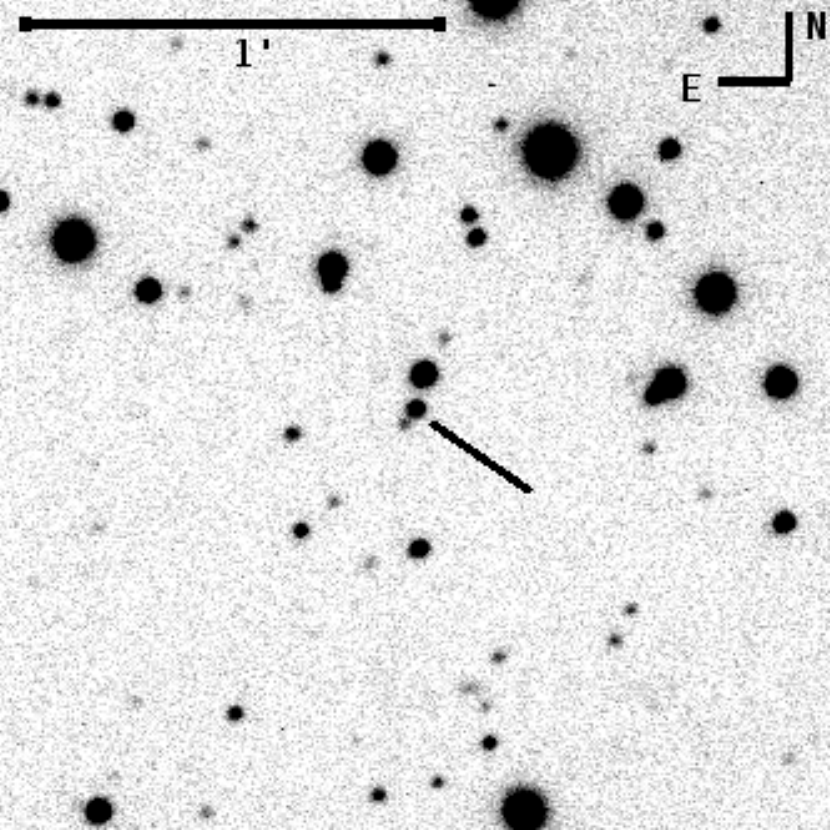}
\end{centering}
\caption{A $2^{\prime}\times2^{\prime}$ finding chart of PTF1J1919+4815 based on a 5\,min exposure with a $g'$ filter obtained using the P200
while the system was in quiescence.
The target is very close to two unrelated stars: a relatively bright star to the northwest
and a very faint star to the southeast. The brighter neighbor is too far away to have a significant impact on
the photometry, while
careful flux extraction and the faint nature of the second neighbor should minimize any contamination.\label{fig:fc}} 
\end{figure}

\begin{figure*}
\includegraphics{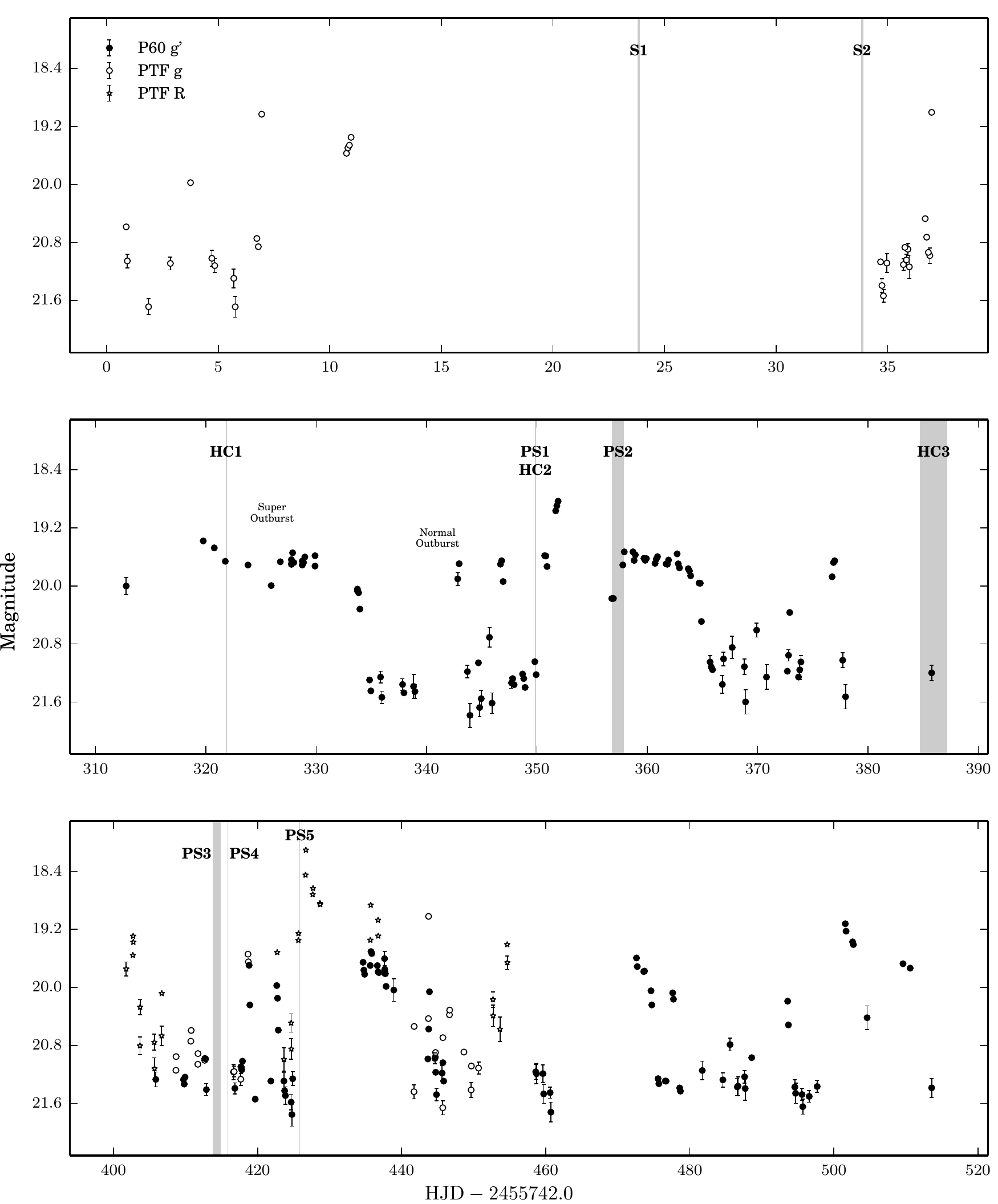}
\caption{A long-term light curve of PTF1J1919+4815, with data from  P60 $g'$, PTF $R$-band, and PTF $g'$-band. Error bars are only shown for those detections with errors of greater than $0.075\,$mag. We label examples of a super outburst and a normal outburst. We also indicate the times of follow-up observations using highlighted areas and labels --- HC indicates high-cadence photometry, S indicates a single spectrum, and PS indicates phase-resolved spectroscopy. More information on the follow-up observations is in Table \ref{tbl:observations}. \newline
The data from the three telescope and filter combinations were not jointly calibrated.
Rather, all data were calibrated against an external source (USNO-B 1.0 $B$-band for $g'$-band observations and $R$-band for $R$ observations) and
for the $g'$-band data sets are very similar (within 10\%). Given the large variability observed here, additional calibration was deemed to be unnecessary.}
\label{fig:longtermlc}
\end{figure*}

Between 2012 May and 2013 April, we obtained both high speed photometry (Section \ref{sec:highcad}) and
phase-resolved spectroscopy (Section \ref{sec:phaseres}).
Simultaneously, we began long-term photometric monitoring of PTF1J1919+4815 (Section \ref{sec:longtermphot}). A summary
of all non-PTF observations
is presented in Table \ref{tbl:observations}. A finding chart, useful given the crowded nature of the field, is shown in Figure \ref{fig:fc}.
A long-term light curve of PTF1J1919+4815, indicating
the times of the higher-cadence observations described below, is presented in Figure \ref{fig:longtermlc}.
Most data presented here are publicly available on the PTF website.

\begin{deluxetable*}{ccccccc}

\tablecaption{Details of Selected Observations\label{tbl:observations}}
\tablecolumns{7}
\tablehead{\colhead{Designation} & \colhead{UT Date} & \colhead{State\tablenotemark{a}} & \colhead{Telescope} & \colhead{Setup} & \colhead{\# of Exp.} & \colhead{Exp. Time (s)}}
\startdata
S1 & 2011 Jul 22 & Outburst & P200/DBSP & B: 600/4000, R: 158/7500 & 1 & 1200\\
S2 & 2011 Aug 02 & Quiescence & P200/DBSP & B: 600/4000, R: 158/7500 & 1 & 900\\
S3 & 2012 Feb 01 & Quiescence & P200/DBSP & B: 600/4000, R: 316/7500 & 1 & 1480\\
HC1 & 2012 May 16 & Outburst & Lick/Kast & Imaging ($g'$) & 201 & 30\\
PS1 & 2012 Jun 13 & Quiescence  & Keck/LRIS & B: 600/4000, R: 600/7500 & B: 81, R: 76 & 120\\
HC2 & 2012 Jun 13 & Quiescence & Lick/Kast & Imaging ($g'$) & 203 & 30\\
PS2 & 2012 Jun 20 & Outburst & GTC/Osiris & R1000B & 40 & 120 \\
\nodata & 2012 Jun 21 & Outburst & GTC/Osiris & R1000B & 21 & 120\\
HC3 & 2012 Jul 18 & Quiescence & Lick/Kast & Imaging ($g'$) & 575 & 15\\
\nodata & \nodata & \nodata & \nodata & \nodata & 100 & 30\\
\nodata & 2012 Jul 19 & Quiescence & Lick/Kast & Imaging ($g'$) & 55 & 30\\
\nodata & 2012 Jul 20 & Quiescence & Lick/Kast & Imaging ($g'$) & 257 & 30\\
PS3 & 2012 Aug 16 & Quiescence & Keck/LRIS & B: 600/4000, R: 600/7500 & B: 75, R: 69 & 120\\
\nodata & 2012 Aug 17 & Outburst & Keck/LRIS & B: 600/4000, R: 600/7500 & B: 84, R: 78 & 120\\
PS4 & 2012 Aug 18 & Quiescence & Gemini/GMOS-N & B600 & 103 & 125\\
PS5 & 2012 Aug 28 & Outburst & Gemini/GMOS-N & B600 & 110 & 125\\
HC4 & 2013 Apr 04 & Outburst & P200/CHIMERA & Imaging ($g'$) & 954 & 5
\enddata
\tablecomments{The telescopes/instruments referenced above are as follows:\\
Gemini/GMOS-N: Gemini North 8-m telescope with the GMOS-N imaging spectrograph \citep{Hook:2004bh}.\\
GTC/OSIRIS: GranTeCan 10.4-m telescope with the OSIRIS imaging spectrograph \citep{Cepa:1998qf}.\\
Keck-I/LRIS: Keck-I 10-m telescope with the Low Resolution Imaging Spectrometer \citep{1995PASP..107..375O,1998SPIE.3355...81M}.\\
Lick/Kast: Shane telescope at the Lick Observatory with the Kast imaging spectrograph \citep{Miller:1988uq,KAST}.\\
P200/DBSP: Palomar $200^{\prime\prime}$ telescope with the Double Spectrograph \citep{1982PASP...94..586O}.\\
P200/CHIMERA: Palomar $200^{\prime\prime}$ with the Caltech High-speed Multi-color CamERA
(Section \ref{sec:photomdata}).}
\tablenotetext{a}{Indicates the photometric state of the system between quiescence and outburst. We base this determination
only on the rough photometric magnitude of the system at the time of observation.}
\end{deluxetable*}

\subsection{High-cadence photometric observations}
\label{sec:highcad}

The high-cadence photometric observations provide the most unambiguous
measurements of PTF1J1919+4815's geometric configuration. 
In 2012 May, June, and July we obtained several series of
exposures with the Lick 3-m Shane telescope using a $g'$ filter and 15--30\,s exposures (see Table \ref{tbl:observations}). The 3-month baseline
of these observations provides the best estimate for a period. However, the relatively long exposure times result
in a poor time resolution relative to the short orbital period of AM CVn systems. Our more recent data set,
from the P200 telescope with the CHIMERA instrument, has similar signal-to-noise per exposure but with only 5\,s integrations
and effectively no dead time between exposures. This light curve is ideal for studying the intra-orbital photometric
variability of PTF1J1919+4815, but its short length of only $\mathord{\sim}1\,$hr precludes its use for a period determination.
We use the CHIMERA data set to show the presence of an eclipse (Section \ref{sec:chimera}) and the Lick data
set to identify the precise orbital period (Section \ref{sec:licklcs}).

\subsubsection{CHIMERA Light Curve}
\label{sec:chimera}

\begin{figure*}
\includegraphics{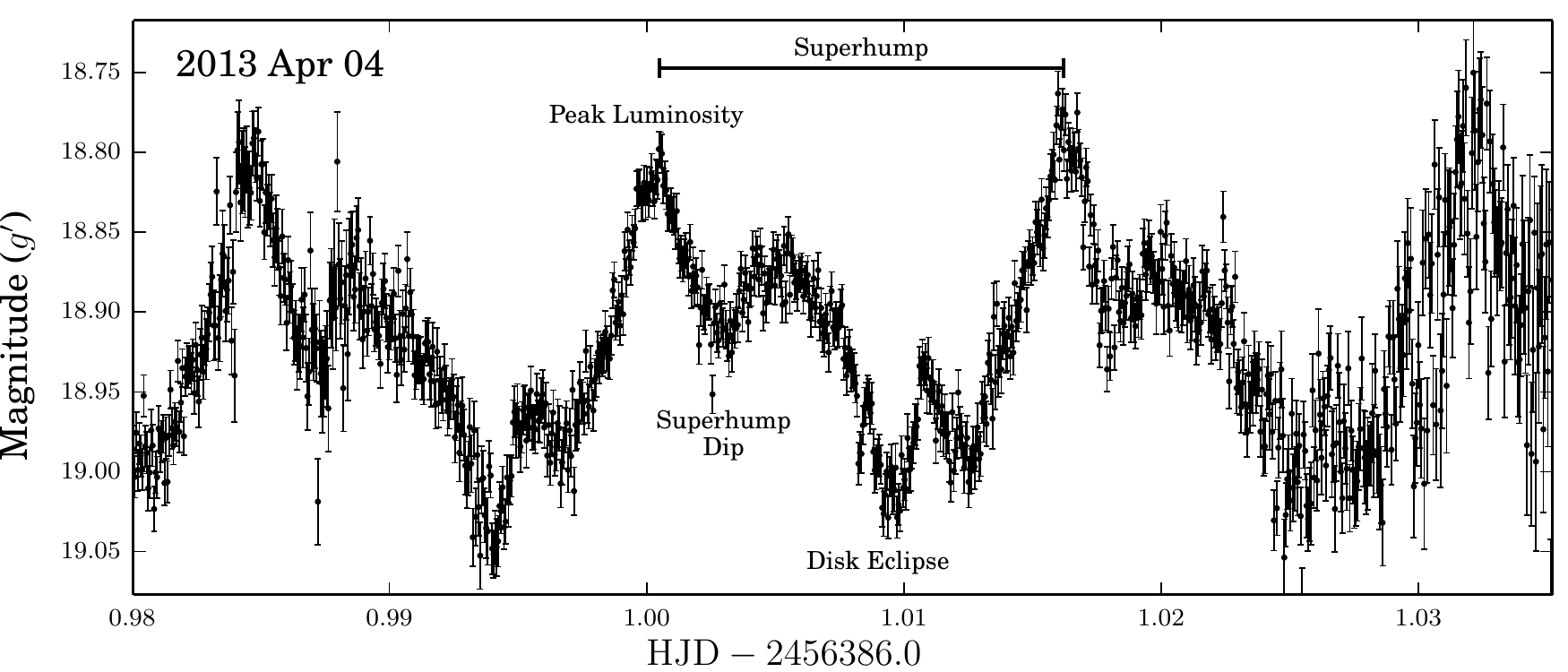}
\caption{The light curve of PTF1J1919+4815 taken using the CHIMERA instrument on 2013 Apr 4. The high
time resolution (5\,s exposures and effectively no dead time between exposures) resolves many features
of the photometric variability; prominent features are labeled. PTF1J1919+4815 was in outburst at the time of observation, and thus shows
a superhump structure as well as the eclipse of the disk. The increased scatter towards the end of the observations
is due to the brightening sky.\label{fig:chimeralc}}
\end{figure*}

We present the CHIMERA light curve in Figure \ref{fig:chimeralc}. At the time of observation, PTF1J1919+4815
was in outburst, and shows the characteristic superhumps seen in other outbursting AM CVn systems
\citep[e.g., Figure 4 of ][]{2002MNRAS.334...87W}. Superhumps are believed to be caused by deformation of the disk
while the system is in outburst, and typically have periods a few percent longer than the orbital period \citep{WARNER1995}.
The most prominent features noticeable besides
the sawtooth shape of the superhumps are the three ``dips'' in luminosity for each orbit of PTF1J1919+4815. 
We now consider the nature of these dips.

We begin by considering the dip which occurs shortly before peak luminosity for each superhump cycle.
If one were to remove the other two dips and draw a straight line from peak luminosity to minimum luminosity,
then this last dip constitutes the minimum of the superhump sawtooth shape. Hence, we do not
believe it to have a geometrical cause beyond the superhump phenomenon itself.

We now turn our attention to the remaining two dips. The presence of one dip with a $\mathord{\leq}5\%$ decrease (labeled 
as ``superhump dip'' in Figure \ref{fig:chimeralc}) has
been observed in other AM CVn systems
\citep[e.g., CR Boo and V803 Cen both show such features; ][]{1997PASP..109.1100P,2000PASP..112..625P}, but its cause is unknown.
However, a second decrease
of any kind has not been observed in other systems except SDSSJ0926+3624 (C11), the first-discovered
eclipsing AM CVn system. Hence, we tentatively conclude that the dip near the peak of
the superhump is a feature intrinsic to the superhump itself, while the second dip is an eclipse of some localized emission on the disk,
possibly the hot spot. This conclusion is supported by additional data we acquired from the Shane telescope,
which we discuss in Section \ref{sec:licklcs}.

The CHIMERA light curve indicates only a single eclipse, with a relatively symmetric ingress and egress. We did consider
the possibility that the third dip, which we earlier identified as part of the sawtooth pattern, is an eclipse of the hot spot, making the
earlier dip an eclipse of the white dwarf. However, the separation of $\mathord{\sim}4\,$min between these two dips means that the
donor would have to eclipse the hot spot at an orbital phase offset of $\mathord{\sim}0.2$, something that is highly unlikely.

In addition to the prominent dips, we also highlight the presence of significant variability on the 1--2\,min timescale,
particularly immediately after the superhump dip.
The source of this variability is unknown and further observations are necessary to ensure that it is, in fact, real.
If so, it may be related to the flickering phenomenon seen in CVs \citep{1992A&A...266..237B}.

\subsubsection{Lick Light Curves}
\label{sec:licklcs}

\begin{figure*}
\includegraphics{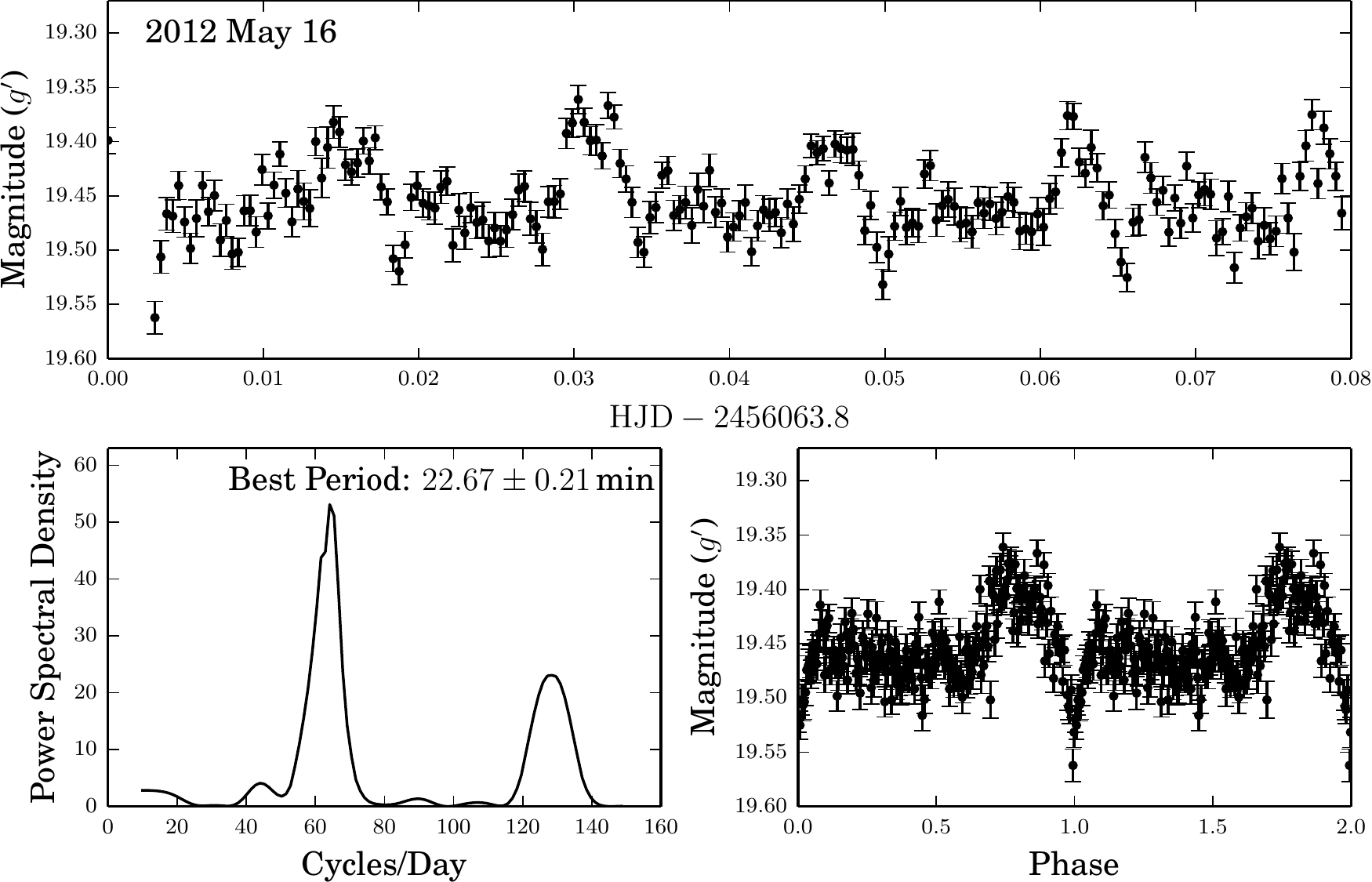}
\caption{A high-cadence light curve of PTF1J1919+4815 observed by the Lick 3-m on 2012 May 16. The top panel is the light curve,
the bottom-left panel is a periodogram of the light curve, and the bottom-right panel is the light curve folded on the orbital period
proposed in Section \ref{sec:orbper}. The eclipse is at a phase of 0.0 using the
ephemeris calculated in the same section. 
The eclipse is clearly visible in the data presented here. It is unknown whether the variability is due to superhumps or the hot spot.}
\label{fig:hclc0516}
\end{figure*}

\begin{figure*}
\includegraphics{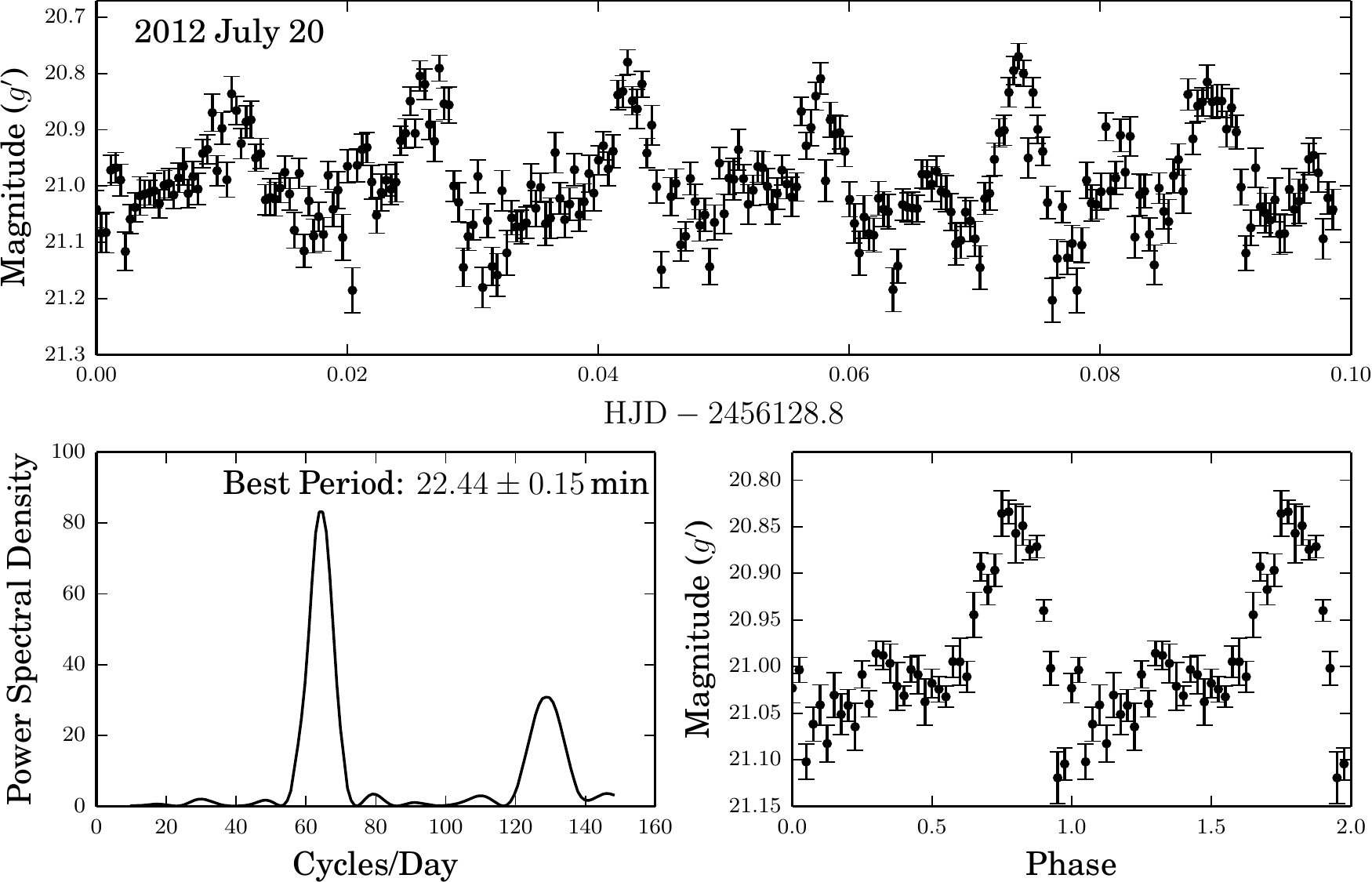}
\caption{A high-cadence light curve of PTF1J1919+4815 observed by the Lick 3-m. This light curve is from 2012 July 20.
The top panel is the light curve, the bottom-left panel is a periodogram generated from the light curve, and the bottom-right panel
is the light curve phased-binned at the orbital period proposed in Section \ref{sec:orbper}. The eclipse is at a phase of 0.0 using the
ephemeris calculated in the same section. This light curve was taken while the system was in quiescence, and shows the quiescent
variability seen in other systems and believed to be directly related to the orbital period \citep{2011ApJ...739...68L}. The data from 
2012 July 18 and 19 are similar.}
\label{fig:hclc0720}
\end{figure*}

The data we obtained from the Lick Shane telescope, while of coarser time resolution, are particularly useful for period
analysis due to its long baseline. We present light curves, Lomb-Scargle periodograms, and folded light curves
in Figures \ref{fig:hclc0516} and \ref{fig:hclc0720} for the nights of 2012 May 16 and 2012 July 20, respectively.
For the data from each night, we subtracted a first order linear fit to remove any
linear trend. The remaining nights show similar characteristics to these, and are both
included in the overall period analysis in Section \ref{sec:orbper}. All light curves are available on the PTF website.

Both the 2012 May and 2012 July data show variability with an amplitude of $\mathord{\sim}0.15\,$mag, including both a
small increase in luminosity as well as the eclipse feature. The eclipse is of similar depth to that seen in the CHIMERA data.
We believe that the luminosity increase in the 2012 July data
is a result of the hotspot rotating into view, as has been shown for CVs \citep[e.g., OY Car; ][]{1983A&A...128...37S} and 
AM CVn systems. \citet{2011ApJ...739...68L} explicitly showed the link between the hot spot and the variability for PTF1J0719+4858, but the
variability in quiescence has been observed in several AM CVn systems \citep[e.g., ][]{1997PASP..109.1100P,2002MNRAS.334...87W}.

The origin of the 2012 May variability is more difficult to determine. PTF1J1919+4815 was in outburst at the time the data were obtained
(see Figure \ref{fig:longtermlc}) and this would typically indicate the presence of superhumps (as seen in the CHIMERA data; see
Section \ref{sec:chimera}). Superhumps, however, typically have a period slightly longer than the orbital period
\citep{2005PASP..117.1204P} and hence should be out of phase with the eclipse. We note that WZ Sge, a CV which also shows
only a disk eclipse, has shown significant differences in eclipse shape between outburst and quiescence \citep{2002PASP..114..721P}.
Given the faintness of PTF1J1919+4815 in quiescence, we do not have sufficient signal-to-noise to clearly see any
such differences in this system.

In this case, the eclipse in the 2012 May data is at the same phase with respect to the luminosity increase
as in the 2012 July data. With fewer than two hours of data, it is impossible to measure the period of the variability with
sufficient precision to distinguish between the orbital period and any slightly longer superhump period. Hence, we cannot say whether
this variability is from the hot spot (thus indicating the lack of superhumps for part of the outburst; see, e.g., C11)
or a chance superposition of the orbital period and the superhump period (as can be seen, for example, in Figure 3 of C11 for
SDSSJ0926+3624).

We now consider the origin of the observed eclipse. The accretor is believed to be significantly more compact than the donor
and contributes the majority of the luminosity of at least some AM CVn systems (e.g., \citealt{2006ApJ...640..466B}; C11). Any eclipse of the
accretor would likely have a sharp ingress and egress and a depth of greater than 10\% (as observed for PTF1J1919+4815). This leaves
only a localized area of emission on the disk as a possible explanation. It is likely that this is the hot spot typically observed
for semi-detached binaries, particularly for the data taken in quiescence. However, we are cautious about concluding that all data,
including that taken in outburst, shows an eclipse of the hot spot as opposed to some other disk feature,
as it is believed that the superhump phenomena results in a somewhat more
extended area of disk emission \citep{1998ApJ...506..360}. 

\subsubsection{Orbital Period Analysis}
\label{sec:orbper}

We stress that while the eclipse feature is certainly not strong, it is stable and repeating over data sets obtained with
two different telescopes and over a time span of over 10 months, as well as in both quiescent and outburst states of
PTF1J1919+4815. We thus conclude that this is an eclipse of the hot spot and proceed with a period analysis.

Outbursting AM CVn systems have been observed to show photometric variability at two periods, one
in outburst (from the superhumps) and one in quiescence \citep[likely from the hot spot; ][]{2011ApJ...739...68L}.
We begin our analysis with the 2012 July data, which, being in quiescence, should show variability at the orbital
period of the system. The best periods obtained from the individual light curves were
$22.41\pm0.09\,$min and $22.44\pm0.11\,$min for
the nights of 2012 July 18 and 2012 July 20, respectively. If we analyze all the 2012 July data simultaneously,
we find a best period of $22.46\pm0.09\,$min. Finally, if we analyze all data from 2012 May, June, and July simultaneously,
we find a best period of $22.4559(3)\,$min. All errors are calculated from bootstrap simulations,
as described in Section \ref{sec:perest}. The times used in all calculations are based on the local time
stamps provided by the Kast instrument, offset to UTC time, and subsequently corrected for light-travel time
by conversion to HJD. We note that the error estimates of the period do not change
linearly with respect to the baseline, as might be expected. This may be due to slight changes in the light curve between
nights, perhaps similar to that seen for WZ Sge \citep{2002PASP..114..721P}. 
Lastly, the closest alias is at 22.4671\,min, which is $37\sigma$ from the best period
and shows no eclipse in a phase-binned light curve. 

The long exposure times used, the weakness of the eclipse, and the faint nature of the system make ephemeris
determination non-trivial. To obtain an estimate with an accurate error estimate, we used a matched filter
technique. For our template, we selected a triangular function of width 3.7\,min and depth 0.1\,mag (roughly
consistent with the shape of the eclipse). We calculated the cross-correlation of the template with the light curve
using a $\Delta t=1\,$s and identified the maxima of this function for each orbital period. We limited the data used
to that from 2012 May 16, the first half of 2013 Jul 18, and 2013 Jul 20. If the remainder of the data is included, the
resulting ephemeris does not change significantly, but the error, calculated based on the scatter of the observed minima
with the predicted minima, is significantly higher due to poorer conditions in that data.

\begin{figure*}
\includegraphics{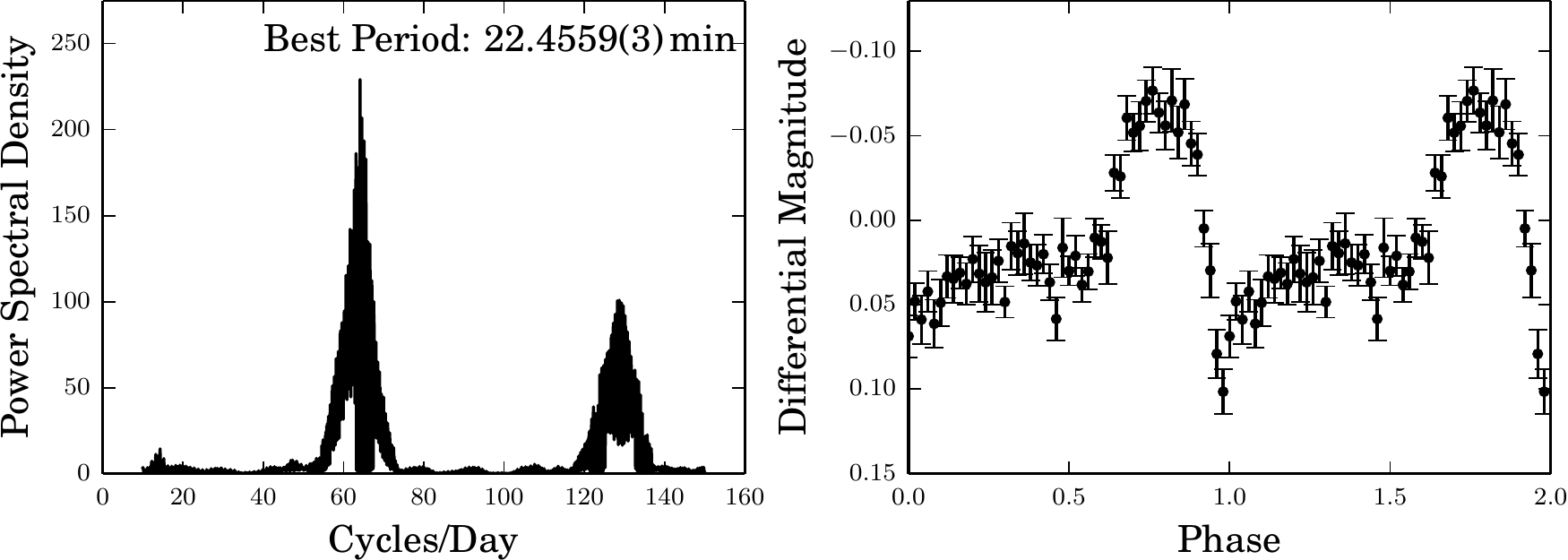}
\caption{A periodogram and phase-binned light curve of all observations of PTF1J1919+4815 on the Lick-3m. The light curve is folded at a
period of 22.4559\,min, the highest peak in the periodogram, and we plot the eclipse at a phase of 0.0 using the ephemeris as described in
Section \ref{sec:orbper}. The error bars on the light curve are the sample standard deviation of the measurements in each phase bin.
\label{fig:pblc}}
\end{figure*}

Given the agreement in the periods measured from these light curves, we propose $22.4559(3)\,$min as the orbital period
of PTF1J1919+4815 and an ephemeris of
\begin{equation*}
HJD=2456063.8650(5)+0.0155944(2)E
\end{equation*}
for the mid-point of the eclipse. We acknowledge that the error measurement of the orbital period may be overly optimistic, as it is from
a bootstrap simulation and does not include systematic errors. However, it is consistent with the long baseline
of observations and when all data are folded at this period, the eclipse is clearly visible (Figure \ref{fig:pblc}).

\subsection{Phase-resolved Spectroscopy}
\label{sec:phaseres}

\begin{figure}
\includegraphics{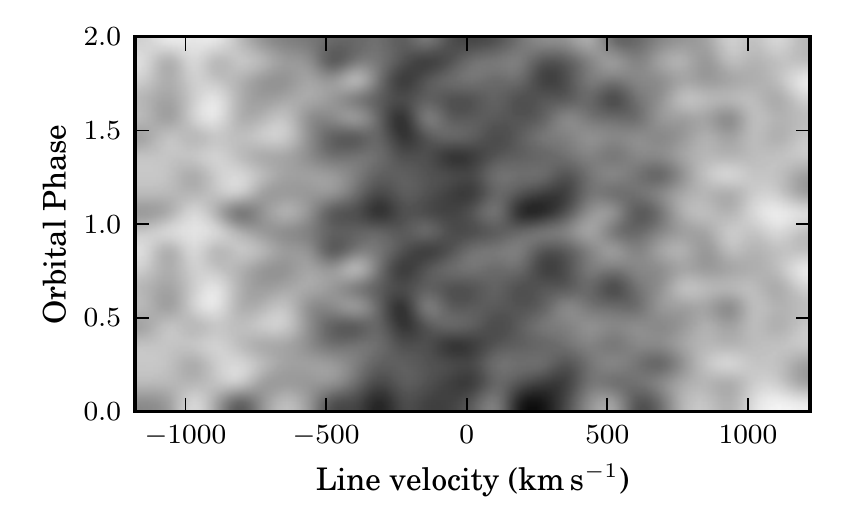}
\includegraphics{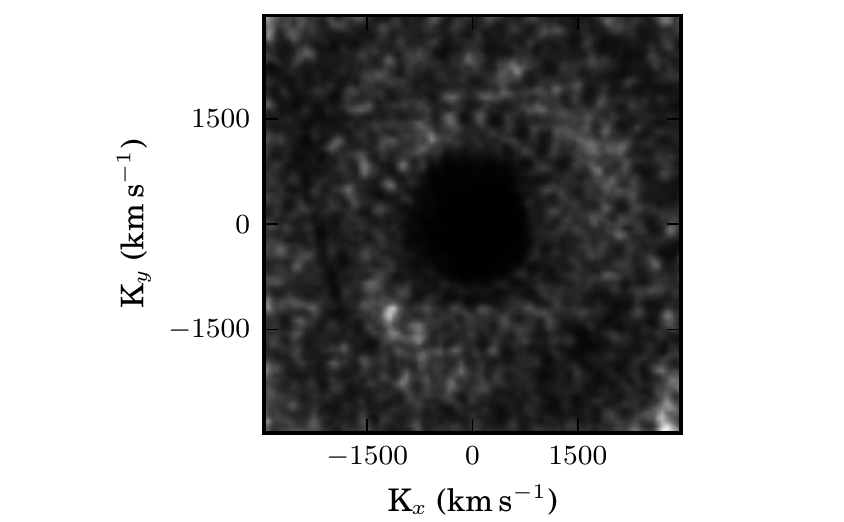}
\caption{\textbf{Top:} The trailed, folded spectra from 2012 June 13 and 2012 August 16 folded at a period of 22.456\,min. The \ion{He}{1} $\lambda\lambda4387,4471,4921,\text{ and } 5015$ lines were used.
Unlike other AM CVn systems, no strong S-wave is present.\\
\textbf{Bottom:} Doppler tomogram generated from data taken on 2012 June 13 using the \ion{He}{1} $\lambda\lambda3888, 4026, 4921, \text{ and } 5015$ lines and the \ion{He}{2} $\lambda4686$ line. Doppler tomograms re-project spectroscopic data into Keplerian-velocity
space --- all flux from areas moving with the same velocity will be projected onto the same spot on the tomogram. This allows us to look for
features such as a hot spot that would be at one velocity. No significant hot spot is seen in this tomogram, unusual for
an AM CVn system.\label{fig:prspec}}
\end{figure}

We obtained phase-resolved spectroscopy on six separate nights, although the outburst state was different for almost
every one of these nights (see Figure \ref{fig:longtermlc} and Table \ref{tbl:observations}). For each night, we co-added
the flux around several He emission lines in velocity space and folded the spectra on the orbital period identified in
Section \ref{sec:orbper}. However, no clear ``S-wave'' was observed, as would be expected for an AM CVn system
\citep{1981ApJ...244..269N}. Likewise, a search for an orbital period in this data did not yield any discernible orbital
period. We present one of these trailed, phase-folded spectra in Figure \ref{fig:prspec}.

Next, we generated Doppler tomograms \citep{1988MNRAS.235..269M} using multiple emission lines (which provides for
better signal to noise in faint, shorter period systems --- see \citealt{2013MNRAS.432.2048K} for an example).
The tomogram should boost the signal of the hot spot since all light is concentrated at one point in velocity space,
yet we did not see any strong hot spot in any of the Doppler tomograms (Figure \ref{fig:prspec}).

The lack of a clear signal in the trailed, phase-folded spectra and the Doppler tomograms is mysterious. We do not
attribute this to a lack of data, as \citet{Levitan:2013uq} found a faint S-wave in a $g'>22$ AM CVn system with a
similar amount of data as what we obtained for PTF1J1919+4815. Time variability of the visibility of the hot spot has
been observed for other AM CVn systems \citep{2009MNRAS.394..367R} and may be the reason we could not identify
an S-wave in the data. We thus conclude that the hot spot in PTF1J1919+4815
had lower contrast with the accretion disk than in other systems during the times it was observed, but
note that this does not change our conclusions about the orbital period, as no clear signal was found in the spectroscopic
data and the photometric signal is much less ambiguous.

\subsection{Long-term Photometric Variability}
\label{sec:longtermphot}

We explore the long term photometric variability of PTF1J1919+4815 using the 355 photometric measurements over $\mathord{\sim}200\,$d.
We identify the presence of super-outbursts and the much shorter normal outbursts. The latter, in particular, appear to last $\mathord{\sim}1\,$d
and were observed to occur 2--3 times between super-outbursts, consistent with CR Boo \citep{2000MNRAS.315..140K} and
PTF1J0719+4858 \citep{2011ApJ...739...68L}, both of which have similar orbital periods as PTF1J1919+4815.

\begin{figure}
\includegraphics{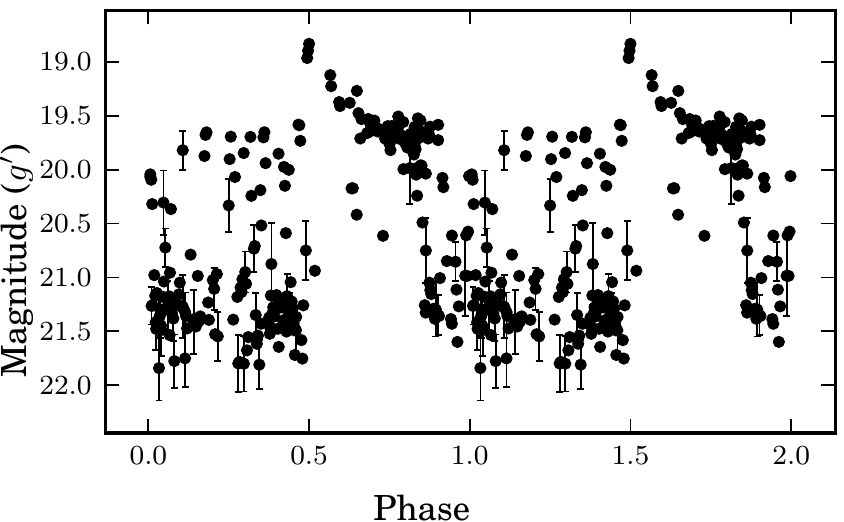}
\caption{A light curve of the long-term P60 photometry folded at a period of $36.8\,$d. The folded
light curve shows a very consistent super-outburst over the 5 super-outburst cycles ($\mathord{\sim}200\,$d) observed.
The normal outbursts, however, do not coincide with each other. Some evidence is present for ``dips'' during the middle of the
super-outburst, as reported by \citep{2012MNRAS.419.2836R}. However, the dips seen here are non-coherent, and it is unknown
if these dips are related to those seen by \citet{2012MNRAS.419.2836R}. The peak of the super-outburst is set to a phase of 0.5.}
\label{fig:longtermanal}
\end{figure}

Our data covers approximately 5 super-outburst cycles and these super outbursts are seen to recur every $\mathord{\sim}35\,$d.
We use a periodogram of the P60 data to calculate the strongest period of $36.8\pm0.8\,$d. Folding the light curve
(see Figure \ref{fig:longtermanal}) at this period results in a remarkably well-defined
super-outburst light curve, something that is not seen in most AM CVn systems.
We note that we use the periodogram itself only to refine the already evident recurrence time.

This super-outburst recurrence time is slightly shorter than that observed for CR Boo, which has an orbital period of 24.5\,min.
The duration of the super-outburst is $\mathord{\sim}13\,$d and its 
amplitude (peak luminosity minus quiescent luminosity) is $\mathord{\sim}3\,$mag.
Although the data shown here indicates regular super-outbursts, long-term study of CR Boo \citep{2013PASP..125..126H} suggests
that this is likely to only be a temporary state. Moreover, with only 5 observed super-outbursts, the scatter in the super-outburst
recurrence time measured here is likely to be underestimated as most AM CVn systems show changes of up to 15\%
(see, e.g., \citealt{2012MNRAS.419.2836R}).

\subsection{Median Spectra and Long-term Variability}
\label{sec:spec}

\begin{figure*}
\includegraphics{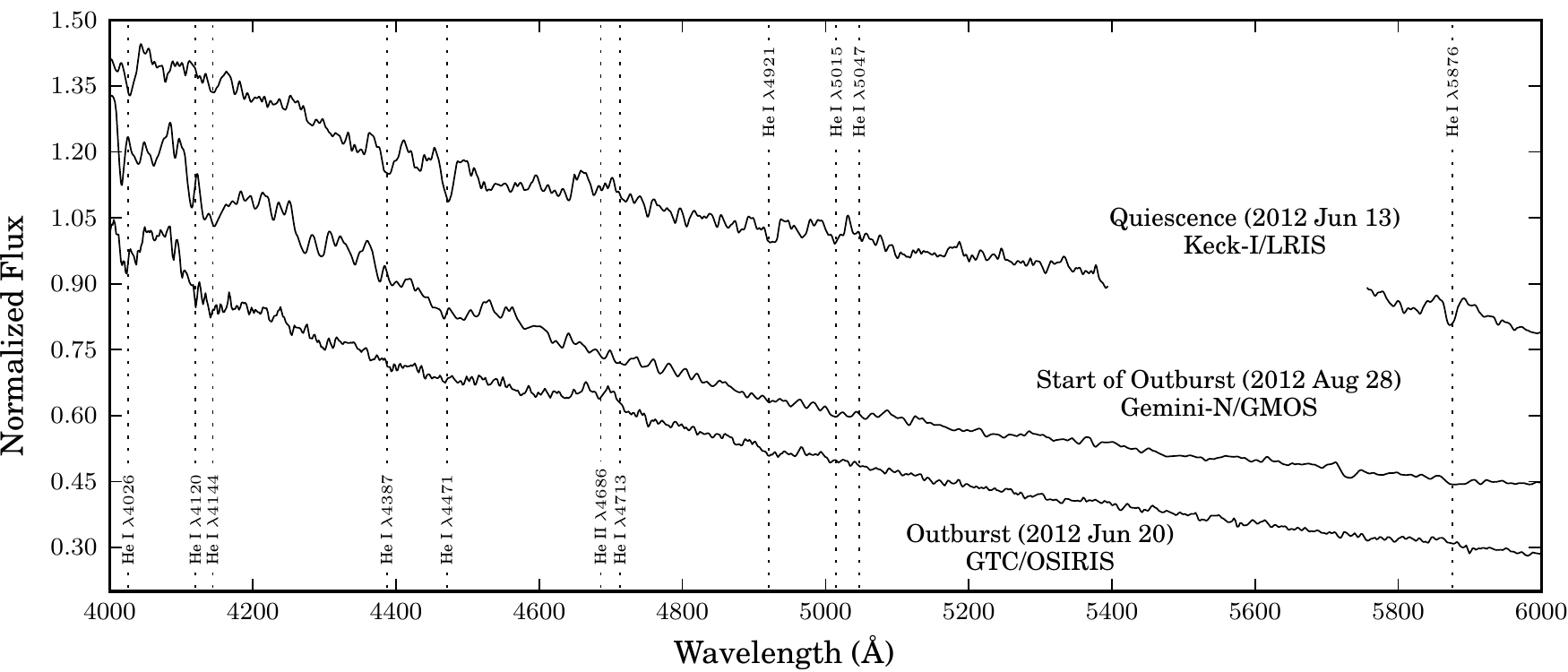}
\caption{The median spectra from three nights: 2012 June 13, 2012 August 28, and 2012 June 13, arranged in this order from top to bottom.
The flux of the first two spectra is shifted, respectively, 0.4 and 0.2 upwards from their normalizations. The most prominent He emission lines
are shown. The first spectrum is of the system in
quiescence and shows the He emission lines that AM CVns are known for. The other spectra are from the outburst state---in particular, the
second spectrum shows few features and is from the initial rise to outburst, while the third spectrum is from the plateau phase and shows the absorption line features that have been observed for other systems in outburst.\label{fig:spectra}}
\end{figure*}

AM CVn systems are known to have significant changes in their spectra corresponding to their different states. PTF1J1919+4815, with its
frequent changes between outburst and quiescence, provides a good opportunity to consider the changes in spectra.
In Figure \ref{fig:spectra} we present median spectra from three nights. In particular, these spectra show
the evolution of the spectrum from the emission-line spectrum that AM CVn systems exhibit in quiescence (the 2012 June 13 spectrum)
to the outburst spectrum showing absorption lines (the 2012 June 20 spectrum). The figure also shows the spectrum taken
on 2012 August 28, at the start of a super-outburst (see Figure \ref{fig:longtermlc}), which shows a distinct lack of features.
We present equivalent widths of the significant lines identified in the outburst and quiescent spectra in Table \ref{tbl:eqwidths}.
\begin{deluxetable}{ccc}
\tablewidth{\columnwidth}
\tablecaption{Equivalent Widths of Prominent Lines\label{tbl:eqwidths}}
\tablehead{\colhead{Line} & \colhead{Quiescence (\AA)} & \colhead{Outburst (\AA)}} \\
\startdata
\ion{He}{1}  3705        &  \nodata\tablenotemark{a} & 4.4 $\pm$ 0.2  \\
\ion{He}{1}  3819        &  -2.7 $\pm$ 0.5          & \multirow{2}{*}{$3.9 \pm 0.2$\tablenotemark{b}}    \\
\ion{He}{1}  3888        &  -4.9 $\pm$ 0.4          &   \\
\ion{He}{1}  3964        &  -1.1 $\pm$ 0.4          & \multirow{2}{*}{$6.8 \pm 0.2$\tablenotemark{c}}  \\
\ion{He}{1}  4026        &  \nodata\tablenotemark{a} &   \\
\ion{He}{1}  4120/4143   &  X                       & 4.4 $\pm$ 0.2  \\
\ion{He}{1}  4388        &  -0.5 $\pm$ 0.4          & 1.5 $\pm$ 0.1  \\
\ion{He}{1}  4471        &  -2.2 $\pm$ 0.4          & 3.3 $\pm$ 0.2  \\
\ion{He}{2} 4685         &  \multirow{2}{*}{-3.4 $\pm$ 0.4\tablenotemark{d}}     &  -0.5  $\pm$ 0.1     \\
\ion{He}{1} 4713         &                          & \nodata\tablenotemark{a}  \\
\ion{He}{1} 4921         &   \nodata\tablenotemark{a} & 2.8 $\pm$ 0.2  \\
\ion{He}{1}  5015        &  -3.6 $\pm$ 0.4          & 1.8 $\pm$ 0.2  \\
\ion{He}{1}  5875        &  -2.7 $\pm$ 0.6          & 1.9 $\pm$ 0.3  \\
\ion{He}{1}  6678        &   2.6 $\pm$ 0.5          &  \nodata\tablenotemark{e}  \\
\ion{He}{1} 7065         &  -2.1 $\pm$ 0.5           & X \\
\enddata
\tablecomments{Quiescent equivalent widths are measured from the medium spectrum taken on 2012 August 16. Outburst
equivalent widths are measured from the median spectrum taken on 2012 August 17. An X indicates that the line is not detectable
above the noise level of the spectrum obtained.}
\tablenotetext{a}{Line present but insufficient S/N to measure.}
\tablenotetext{b}{Combined equivalent width of \ion{He}{1} 3819 and \ion{He}{1} 3888.}
\tablenotetext{c}{Combined equivalent width of \ion{He}{1} 3964 and \ion{He}{1} 4026.}
\tablenotetext{d}{Combined equivalent width of \ion{He}{2} 4685 and \ion{He}{1} 4713.}
\tablenotetext{e}{Line present but contaminated with atmosphere.}
\end{deluxetable} 

\section{Discussion}
\label{sec:discussion}

Although the presence of the disk eclipse allows for a precise period estimate for such a faint system,
the lack of an accretor eclipse precludes a full solution of the system structure.
It also prevents us from measuring any change in the orbital period for this system, as the uncertainty on the disk
radius and its likely change due to the presence of outbursts will overwhelm any expected change due to orbital evolution.
However, the presence of the disk eclipse does allow us to estimate the inclination of the system if we make certain assumptions
about the masses and radii of the two components. 

Only one AM CVn system, SDSSJ0926+3624 ($P_{orb}=28.558\,$min; C11), has been observed to have eclipses of the accretor: 
and as a result, has had a full system solution derived.
Here, we assume that the evolutionary origin and, hence, system properties of PTF1J1919+4815
are similar to that of SDSSJ0926+3624. We caution that AM CVn systems
are believed to have several possible evolutionary pathways \citep{2001A&A...368..939N} and that no conclusive
evidence has been found that can differentiate a system's origin \citep{2007ApJ...666.1174R,2010MNRAS.401.1347N}.

From C11, the parameters of SDSSJ0926+3624 are: $M_1=0.85\msol$, $M_2=0.035\msol$,
$R_1 = 9.7 \times10^{-3}\rsol$, and $q=M_2/M_1=0.041$. We use these values for PTF1J1919+4815, despite the difference in
orbital period. Assuming standard mass transfer, only $\mathord{\sim}0.01\msol$ would have been transferred during the
evolution between the orbital periods of PTF1J1919+4815 and SDSSJ0926+3624, which is certainly below the uncertainties
of the systems' respective evolutionary histories. Similarly, we adopt the radius of the
accretor from SDSSJ0926+3624 to be that of the accretor in PTF1J1919+4815. The donor's radius will be that of the Roche
Lobe as a first estimate, at the given period for the given mass ratio. Using the assumed masses for PTF1J1919+4815
and its orbital period with the approximation in \citet{Eggleton:1983fk}, we find that
the orbital separation, $a$, and the Roche-lobe radius, $r_L$, are $a=0.25\rsol$, and $r_L=0.039\rsol$. The last measurement needed is the radius of the disk. Based on the measurements of
SDSSJ0926+2624, we assume a disk radius of $R_{disk}\approx0.35a$. The derived values in C11
vary between observing runs, and since the size of the disk is expected to change between outburst and quiescence, we believe
this is a reasonable, first assumption. We can constrain the inclination by determining for which inclinations the shadow of the
donor star (as seen from Earth) passes over the hot spot, assumed to be a point source, but not the accretor. Combining the
radii of disk, donor, and accretor, we can constrain the inclination to $76^{\circ} < i < 79^{\circ}$. We present a summary
of system properties in Table \ref{tbl:props}.

Given the uncertainty in the system properties used, we consider the impact of several parameters on the results.
The upper limit on $i$ is primarily related to the accretor and donor masses.
However, their masses are constrained by the initial conditions of AM CVn system formation and
as such are not likely to change significantly even with moderate changes of the initial masses.
The lower bound, however, is significantly affected by the radius of the disk. If, for
example, the radius of the disk was $0.5a$, the lower bound would drop to $72^{\circ}$. Had the inclination been greater than the
upper limit here, an eclipse of the white dwarf should have been observable. Conversely, an inclination of less than the lower limit
would result in no eclipse of either the hot spot or the accretor.

\begin{deluxetable}{lc}
\tablecaption{System Properties\label{tbl:props}}
\tablecolumns{2}
\startdata
\tableline\\
$U$\tablenotemark{a} & $19.814\pm0.048$\\
$g$\tablenotemark{a}  & $20.155\pm0.016$\\
$r$\tablenotemark{a}  & $20.522\pm0.065$\\
$i$\tablenotemark{a}  & $20.409\pm0.121$\\
\\
Orbital period & $22.4559(3)$\,min\\
Super-outburst Recurrence Time & $36.8\pm0.4\,$d\\
Super-outburst Strength & $\mathord{\sim}3\,$mag\\
Super-outburst Duration & $\mathord{\sim}13\,$d\\
Inclination\tablenotemark{b} & $74^{\circ} < i < 78^{\circ}$
\enddata
\tablenotetext{a}{Photometric measurements are from the Kepler INT Survey DR2 \citep{Greiss:2012ly,Greiss:2012zr} and are
typical of the quiescent values for the system.}
\tablenotetext{b}{These values assume that the radius of the accretion disk, $R_{disk}=0.35a$, where $a=0.25R_\odot$. See Section \ref{sec:discussion} for further details.}
\end{deluxetable}

\subsubsection{Number of Eclipsing AM CVn Systems}

One peculiarity of the currently known population of AM CVn systems is the relative scarcity of eclipsing examples. In Section \ref{sec:discussion}, we constrained the
inclination of PTF1J1919+4815 to the range $76^{\circ} < i < 79^{\circ}$. This indicates that in $\mathord{\sim}19\%$ of systems with
similar orbital periods the accretor should be eclipsed, and an additional $\mathord{\sim}5\%$ of similar systems should show
an eclipse of the hot spot only, as is the case
here. These calculations assume a random distribution of inclinations and an observed inclination distribution of $\sin(i)$.
We note that longer orbital period systems will have a smaller $q$ and hence a lower probability of eclipse.

With the number of AM CVn systems rapidly approaching 40, it is somewhat surprising that only one system has shown an
eclipse of the accretor and only one other a partial eclipse. The rate of observed eclipsing systems, $\mathord{\sim}5\%$ is
lower than would be expected, even assuming that only $15\%$ of systems are eclipsing (based on $q=0.02$ for longer period 
systems). Even accounting for small-number statistics, one would expect at least 3 fully eclipsing systems.
We do note that many long period systems do not have high-cadence photometric follow-up, but only time-resolved spectroscopy. However, a prominent eclipse should also show up in time-resolved spectroscopy through flux and/or line-strength/shape changes.

\section{Conclusions}
\label{sec:conclusions}

We have established PTF1J1919+4815 as an eclipsing AM CVn system from its He-rich and H-deficient spectrum, its phenomenological behavior,
and its photometrically determined period. Its orbital period was measured to be $22.4559(3)\,$min,
but we did not find a significant hot spot in the S-wave and Doppler map. PTF1J1919+4815
shows a well-defined super-outburst that is $36.8\pm0.4\,$d. Such a non-variable recurrence time is atypical
for AM CVn systems. We used these measurements to constrain the inclination of the systems given assumptions
about its evolution and to consider the geometric structure of the system.

\acknowledgements

T.~K. acknowledges support by the Netherlands Research School of Astronomy (NOVA).
We thank Dong Xu for reducing the initial classification spectra.

Observations obtained with the Samuel Oschin Telescope at the Palomar
Observatory as part of the Palomar Transient Factory project, a scientific
collaboration between the California Institute of Technology, Columbia
University, Las Cumbres Observatory, the Lawrence Berkeley National
Laboratory, the National Energy Research Scientific Computing Center,
the University of Oxford, and the Weizmann Institute of Science. Some
of the data presented herein were obtained at the W.M. Keck Observatory,
which is operated as a scientific partnership among the California
Institute of Technology, the University of California and the National
Aeronautics and Space Administration. The Observatory was made possible
by the generous financial support of the W.M. Keck Foundation. 
The authors wish to recognize and acknowledge the very significant cultural role
and reverence that the summit of Mauna Kea has always had within the indigenous
Hawaiian community.  We are most fortunate to have the opportunity to conduct observations from this mountain. 
Based in part on observations obtained at the Gemini Observatory, which is operated by the 
    Association of Universities for Research in Astronomy, Inc., under a cooperative agreement 
    with the NSF on behalf of the Gemini partnership: the National Science Foundation 
    (United States), the National Research Council (Canada), CONICYT (Chile), the Australian 
    Research Council (Australia), Minist\'{e}rio da Ci\^{e}ncia, Tecnologia e Inova\c{c}\~{a}o 
    (Brazil) and Ministerio de Ciencia, Tecnolog\'{i}a e Innovaci\'{o}n Productiva (Argentina).
 The Gemini data were obtained under Program ID GN-2012B-Q-110.
Based in part on observations made with the Gran Telescopio Canarias (GTC)
installed in the Spanish Observatorio del Roque de los Muchachos of the Instituto de Astrof\'{i}sica de Canarias, 
on the island of La Palma.

\textit{Facilities:}
\facility{PO:1.2m}, \facility{PO:1.5m}, \facility{Gemini:Gillett (GMOS-N)}, \facility{GTC (OSIRIS)}, \facility{Hale (CHIMERA)}, \facility{Hale (DBSP)}, \facility{Keck:I (LRIS)}, \facility{Shane (Kast Double spectrograph)}\\\

\bibliographystyle{/Users/david/Astro/Papers/apj.bst}
\bibliography{ms}

\end{document}